\documentstyle[preprint,aps,epsf]{revtex}

\begin{document}

\title{
\hfill \parbox[t]{2 in} {\rm \small hep-ph/9511422}
\vskip 1.3 cm
Glueball Spectroscopy in a
	Relativistic Many-Body Approach to Hadron
         Structure.}
    \author{
     Adam Szczepaniak, Eric S. Swanson, Chueng-Ryong Ji, and Stephen R.
Cotanch}
     \address{
     Department of Physics,
      North Carolina State University,
     Raleigh, North Carolina 27695-8202}

       \maketitle

\begin{abstract}

	A comprehensive, relativistic many-body approach to hadron
structure is advanced based on the Coulomb gauge QCD Hamiltonian.
 Our method incorporates standard many-body techniques
which render the approximations amenable to systematic
improvement. Using BCS variational methods, dynamic chiral symmetry breaking
naturally emerges and both quarks and gluons acquire constituent masses.
 Gluonia are studied both in the valence and in the collective, random phase
 approximations.
  Using representative values for the
strong coupling constant and string tension, calculated
  quenched glueball masses are found to be in remarkable agreement
with lattice gauge theory.

\end{abstract}
 \date{November, 1995}
\narrowtext

 Our knowledge  of the standard model cannot  be
considered complete until
 explicit  gluonic degrees of freedom are found and understood\cite{AC}.
 In an effort to address this issue we advance
a  comprehensive framework for consistently describing and
 understanding hadron structure -- including the glueball and hybrid
sectors.  The model is  motivated in part by our previous studies of
relativistic~\cite{NCSU} and
nonrelativistic \cite{NR} quark models and by the pioneering work
of Le Yaouanc  {\it et al.}\cite{O}, who, following the spirit of Nambu and
 Jona-Lasinio\cite{NJL}, have constructed a quark-based model of the QCD
vacuum.
 Only a brief description of our approach
is provided here. A full, detailed treatment will be contained in a
 separate communication.

 The idea is to build on  the known successes of
the constituent quark model for heavy quarks by considering a
 many-body relativistic Hamiltonian in a quasiparticle basis where
  dynamical chiral symmetry breaking and  massive gluon modes are explicit.
 Such a model incorporates an extensive Fock space but reduces to the simple
quark model in the valence approximation. Furthermore,
  the simultaneous presence
of quark and gluon degrees of freedom permits
 studying their mixture in hybrid and glueball
states. This is especially important since glueball
searches tend
to occur in meson-rich regions of the hadron spectrum and also
 because it may be years before lattice gauge calculations provide significant
 insight.
This letter focuses on the gluonic sector of the model Hamiltonian, presenting
the glueball spectrum calculation and a discussion of the
associated approximation schemes. In the summary we  comment on other issues
regarding  applications to mesons, baryons and hybrids.

 There have been a variety of previous glueball studies: the Bag
Model~\cite{JJ,BCM,CHP},
QCD Sum Rules~\cite{S,PT,N,DP}, the Constituent Glue Model~\cite{B},
  and the Flux Tube
Model~\cite{IP}.  These approaches differ markedly in their mass
predictions, sometimes by as much as $1\mbox{ GeV}$, and no single approach has
consistently reproduced lattice gauge measurements~\cite{UKQCD,GF112,T}.
  As stated above, our goal is to model QCD in
 a way which is in accord with the successes of the constituent quark model.
  Therefore we start from the Coulomb gauge QCD Hamiltonian $H$ \cite{L} and
  assume
  the existence of a set of phenomenological interactions
  $H_{phen}$ such that $H$ can be written as
  \begin{equation}
  H = H_0  + H_I \label{H}
  \end{equation}
 where  $H_0$  is defined as
  \begin{equation}
  H_0 = K + H_{phen}, \label{H0}
  \end{equation}
and $K$ stands for the kinetic energy
    \begin{equation}
  K = \int d{\bf x} \Psi_q^{\dag}({\bf x})\left[ -i{\vec \alpha} \cdot{\vec
\nabla}
  + \beta m_q \right]
\Psi_q({\bf x})  + {1\over 2} \int d{\bf x} \left[ |{\bf E}^a({\bf x})|^2
+  |{\bf B}^a(x) |^2  \right],
  \end{equation}
involving current quarks having masses $m_q$ and zero mass transverse gluons.
The residual  potential $H_I = H_I^{QCD} - H_{phen}$ is given by the difference
between the original
  QCD and phenomenological interactions.
  The motivation for introducing phenomenological interactions  is to generate
a much weaker
 residual potential
  $H_I$ at all energy scales.
Furthermore, as in the phenomenological quark model, we will assume that
in the quasiparticle quark and gluon basis this residual
interaction can  be approximated by the canonical QCD interaction Hamiltonian
with a coupling $g_s$ that is small and saturates at low energies. Under this
approximation,  to any   order in $g_s$  the residual  interaction can  in
principle  be derived  using standard
 methods of time independent perturbation theory. However,  since the infrared
 behavior  has  already been determined  from  phenomenology, the residual
 interaction must be free from infrared divergences to avoid double
 counting.
 This can be achieved,  for example, by  imposing a cutoff, $\Lambda_{IR}$,
 on $H_I$ that
 removes coupling
  between quasiparticle states  whose energy difference is smaller than the
  cutoff scale.
 The perturbative expansion of $H_I$ then generates an effective Hamiltonian
 which has
 nonvanishing matrix elements below the cutoff  and can either be added to
$H_0$ and diagonalized nonperturbatively or,  because of the small coupling,
included
 in the bound state perturbation theory  using the  eigenstates of
 $H_0$~\cite{WI}.

    For the applications considered in this work the phenomenological
   Hamiltonian  is taken to be

 \begin{equation}
 H_{phen} =  -{1\over 2}\int d{\bf x} d{\bf y} \rho^a({\bf x})V_L(|{\bf x} -
 {\bf y}|) \rho^a ({\bf y}),
   \label{HL}
 \end{equation}
with color charge density $\rho^a({\bf x}) = \Psi_q^{\dag}({\bf x})T^a
\Psi_q({\bf x})
 + f^{abc} {\bf A}^b({\bf x}){\bf E}^c({\bf x})$;  $V_L$ is
 a  linear confining potential,
   \begin{equation}
   V_L(|{\bf x} - {\bf y}|) =  {2 {N_c b} \over {N_c - 1}} |{\bf x} - {\bf
y}|\,
   \left(1 - {\rm e}^{-\Lambda_{phen} \vert {\bf x} - {\bf y} \vert} \right)
\end{equation}
and  $N_c = 3$ is the number of colors.
The string tension will
 be fixed at 0.18 GeV$^2$,
 commensurate with Regge phenomenology and the naive quark model. Since this
 interaction is meant to represent soft physics, we have introduced an
 ultraviolet cutoff directly into the potential. Note that the approach
 advocated here
 requires that $\Lambda_{phen} \sim \Lambda_{IR}$ and hence to completely
 describe low energy phenomena
 we must add
the QCD interactions below the cutoff $\Lambda_{phen}$ to this phenomenological
Hamiltonian.
To order $\alpha_s$  we only keep the Coulomb potential and ignore
 self-energies,
hyperfine interactions and vacuum corrections. This
will be discussed in more detail below.  The complete interaction in the
Hamiltonian $H$,  which we
diagonalize nonperturbatively is thus given by  Eq.~(\ref{HL}) with $V_L$
replaced by
\begin{equation}
V_L(r) \to V(r) = V_L(r) + V_C(r),
\end{equation}
where
\begin{equation}
V_C(|{\bf x} - {\bf y}|) =  -{\alpha_s \over {  |{\bf x} - {\bf y}| } }
\end{equation}
is the Coulomb potential. Finally we note that both $V_C$ and $H_I$ need to
be ultraviolet regulated. The standard perturbative renormalization procedure
 may then be followed for these terms \cite{AD}.

 Regarding the structure of $H_{phen}$ we have assumed that
 the bulk of the low energy dynamics of the  $q\bar q$ and gluon-gluon systems
 may be described by a two-body interaction as shown and have employed a
linear confinement potential in keeping with the
constituent quark model and lattice
gauge theory.  Also note that the confining interaction
is between color densities
  rather than scalar currents as is usually assumed in the constituent
quark model. This is discussed below.
 Note that Eq.(\ref{HL}) implies that gluons may be
confined into gluon-gluon bound states which forms the basis of our
glueball investigation. However
lattice gauge
results~\cite{lgt} indicate that static color octets become screened
at very large distances and  Eq.(\ref{HL}) does not reflect this.
Physically,
one expects that the screening of the gluonic confinement potential is due to
the
creation of low-lying glueballs at large valence gluon separation. Thus the
model as
presented
here is similar in spirit to the naive quark model, where linear confinement in
the
$q \bar q$ sector is
absolute and it is mixing to other Fock sectors which is responsible for the
screening. Furthermore, we note that constituent
gluons are not static, low lying glueballs tend to be compact, and that lattice
gauge calculations support the picture of a diffuse spectrum.
Finally, one should not think of the confinement potential as a flux tube
because this
quickly leads to conundrums about double counting gluonic degrees of freedom.
The Coulomb gauge Hamiltonian makes a clear distinction between the gluonic
color density and the Coulomb interaction -- one which applies equally
well to the confinement interaction.

As in Ref.~\cite{O} the Fock space in which we diagonalize
$H_0$ is constructed from the variational, BCS vacuum $|\Omega\rangle$
 by an
application of the constituent quark (antiquark) $B^{\dag}$ ($D^{\dag}$)
 and gluon ${\bf a}^{\dag}$ creation operators with

 \begin{eqnarray}
\Psi_q({\bf x}) & = & \sum_{\lambda}\int {{d{\bf k}}\over {(2\pi)^3}}
 \left[U({\bf k},\lambda)B({\bf k},\lambda)
  + V(-{\bf k}, \lambda) D^{\dag}(-{\bf k},\lambda)\right] e^{i{\bf k}\cdot{\bf
x}}
 \nonumber \\
{\bf A}^a({\bf x}) & = &  \int {{d{\bf k}}\over {(2\pi)^3}}
{ 1\over {\sqrt{2\omega(|{\bf k}|)}}} \left[ {\bf a}^a({\bf k}) + {\bf
a}^a({-\bf k})^{\dag}
 \right] e^{i{\bf k}\cdot{\bf x}}
 \end{eqnarray}
 and with the vacuum defined by $B|\Omega\rangle = D|\Omega\rangle =
  {\bf a}|\Omega\rangle = 0$.
We note that the gluon operators are transverse so that one has

\begin{equation}
 [a_i^a({\bf k}), a_j^{b}({\bf q})^{\dag} ] = \delta_{ab} \,(2\pi)^3\delta({\bf
k} - {\bf q}
 )\,
  \left(\delta_{ij} - {\hat k}_i {\hat k}_j  \right)
  \label{trans}
\end{equation}

\noindent
The extra term in the final factor complicates the calculations of the glueball
spectrum
  but is crucial to maintaining the correct gluonic
degrees of freedom.

  The variational gap functions $U$, $V$ and $\omega$ are obtained by
minimizing the ground-state
(vacuum) energy $E_0= \langle \Omega|H_0| \Omega  \rangle $.
This leads to gap equations
 which are equivalent to the Schwinger-Dyson equations for the
quark or gluon  self-energies in the rainbow approximation. For the gluon
spectral function $\omega$ the gap equation is given by

\begin{equation}
\omega(q)^2 \equiv \omega(|{\bf q}|)^2 = q^2 + {N_c \over 4} \int{d {\bf k}
   \over (2 \pi)^3} {\tilde V}(
  {\bf k} + {\bf q}) \left( 1 +
   ({\hat {\bf k}}\cdot{\hat {\bf q}})^2
  \right) { \omega(|{\bf k}|)^2 -
  \omega(|{\bf q}|)^2 \over \omega(|{\bf k}|)} \label{ggap}
\end{equation}

\noindent
A similar variational treatment of the Hamiltonian in the quark sector
 results in the well known realization of dynamical chiral symmetry breaking
by the BCS vacuum.

 The presence of the Coulomb term in $V$ introduces a quadratic cutoff
 dependence,
which can be removed by including the neglected terms in $H_I$  (in particular
 the self-energy corrections  resulting from the expansion of the
 Faddeev-Popov determinant
 and the transverse gluon exchange calculated to order $\alpha_s$).
However, the net effect of the order $\alpha_s$ terms
   in $H_I$  is expected to be small and we simply ignore them when solving
   the gap equation.
 Further investigations are in progress.

A good fit to the numerical solution of Eq.~(\ref{ggap}) is obtained
with

\begin{equation}
  \omega(k) = \sqrt{ k^2 + m_g^2 {\rm e}^{-k^2/\kappa^2}}.
\end{equation}
\noindent
If one defines the gluon mass in terms
of the effective mass as $m_g = m_g(0)$ where $m_g(k) = \sqrt{\omega(k)^2 -
k^2}$ then
the proceeding fit yields $m_g = 0.8$ GeV (and $\kappa = 13$ GeV).

The gluon condensate may be simply calculated within the context of the pairing
ansatz. The result is

\begin{equation}
  \langle {\alpha \over \pi} G^{\mu\nu}_a G^a_{\mu\nu} \rangle
  = {N_c^2 - 1 \over \pi^3} \int_0^\infty dk \, k^2 \, \alpha_s(k) {\left(
  \omega(k) - k \right)^2 \over \omega(k)}.
\end{equation}

\noindent
We have allowed for the possibility that $\alpha_s$ runs above the cutoff
 although
this is not crucial.
The calculated condensate agrees well with the QCD sum rule value
of 0.012 GeV$^4$~\cite{SVZ}
 and is only weakly sensitive to the cutoff above $\Lambda_{phen} \sim 4$ GeV.

	Just as in conventional nuclear structure theory, our BCS
many-body vacuum state can be systematically improved by utilizing
the Tamm-Dancoff (TDA), random phase (RPA), or even more
accurate approximations involving exact diagonalization in an
extensive multiparticle--hole model space.
In the glueball case we have performed both TDA and RPA calculations, however,
 because of the large constituent gluon mass we expect the
Tamm-Dancoff approximation to be a reasonable one. Indeed the $0^{++}$
glueball mass is shifted by less than 2\% in going to the random phase
approximation.
In the glueball rest frame, the  TDA  gluon-gluon bound states are given by

\begin{equation}
  \vert J^{PC} \rangle = \int {d{\bf p} \over {(2\pi)^3}}
    \, \chi^{JPC}_{ij}
({\bf p})\, a_i^{b}({\bf p})^{\dag} a_j^{b}(-{\bf p})^{\dag} \vert
\Omega \rangle
\end{equation}

\noindent
with the glueball wave function $\chi^{JPC}_{ij}$ satisfying

\begin{eqnarray}
  & & E^{JPC} \,{\cal G}^{JPC}_{ij}\, \chi^{JPC}_{ij}({\bf q}) =
   - {N_c\over 4} \int {d{\bf k}\over (2\pi)^3}
      \tilde V(| {\bf k} + {\bf q}|) {(\omega(q) + \omega(k))^2 \over
      2\omega(q)\omega(k)}
	 {\cal F}^{JPC}_{ij}\, ({\bf k},{\bf q})\chi^{JPC}_{ij}({\bf k})
\nonumber \\
 +  & &
\left( \left( \omega(q) + {q^2\over \omega(q)} \right) {\cal G}^{JPC}_{ij}
  + {N_c \over 4}  \int {d{\bf k}\over
(2\pi)^3}
   \tilde V(|{\bf k} + {\bf q}|) {\omega(q)^2 + \omega(k)^2 \over
\omega(q)\omega(k)}\,
   {\cal F}^{JPC}_{ij}({\bf k},{\bf q}) \right) \chi^{JPC}_{ij}({\bf q}).
\label{beqn}
\end{eqnarray}
Here the ${\cal F}^{JPC}_{ij}$'s are determined from coupling the two
transverse
gluons labeled by the Cartesian indices $i,j$ to a state with
 total angular momentum $J$, parity $P$ and charge conjugation
 $C$.
 For example

\begin{eqnarray}
  {\cal F}^{0++}_{ij}({\bf k},{\bf q}) &=& \left(
   1 + ({\hat {\bf k}}\cdot{\hat {\bf q}})^2 \right) \delta_{ij}\\
  {\cal F}^{0-+}_{ij}({\bf k},{\bf q}) &=& {\hat k}_i {\hat q}_j ,
\end{eqnarray}

\noindent
with more complicated expressions for $J \ge 2$.
The functions ${\cal G}^{JPC}_{ij}$ are normalization matrices which arise
from  mixing between different $LS$ states induced by the transverse nature of
the gluon.
We use the Coulomb and linear contributions to $\tilde V$ in Eq. (14).
There are
no Faddeev-Popov terms and transverse gluon exchange is treated as a
perturbation (it is
of order $1/m_g^2$). We note that it is not possible to make a two-gluon
$J=1$ state as is consistent with
Yang's theorem. Such spurious states exist in models with explicitly massive
gluons.

There is an interesting property associated with divergences in the bound state
 equation.
The linear potential is infrared divergent; however, this
potentially problematic divergence is cancelled by the self-energy term
in the kinetic energy. This cancellation appears to be a property of a
density-density interaction. For example, the cancellation does not occur for
scalar quark currents (and indeed, a stable vacuum cannot be obtained with an
interaction between scalar currents). Furthermore, the cancellation appears
to only occur in bound state equations for color singlet objects. This has
been observed
previously in the context of the Bethe-Salpeter approximation to the
$q \bar q $ bound
state problem~\cite{AD}.

The spectrum which results from numerically solving Eq. (\ref{beqn}) is
presented
in Fig.~1 along with recent results from lattice gauge calculations. The
agreement
is remarkably good, especially when it is recalled that the model has been
completely fixed from $q\bar q$ phenomenology.  The spectrum
corresponds to
  the ansatz for a running  $\alpha_s(k)$  from
Ref.~\cite{GI}. Employing a fixed value for the strong coupling,
 $\alpha_s = 0.4$, produced no significant
difference.
Furthermore,  the spectrum is essentially independent of the
 cutoff for $\Lambda_{phen} \gtrsim 4$ GeV.
We conclude that the
model captures the essential features of glueballs. To our knowledge, this
is the only  model of gluonia which successfully reproduces lattice data and
therefore it may provide important insight into glueball structure.

\hbox to \hsize{%
\begin{minipage}[t]{\hsize}
\begin{figure}
\epsfxsize=4in
\hbox to \hsize{\hss\epsffile{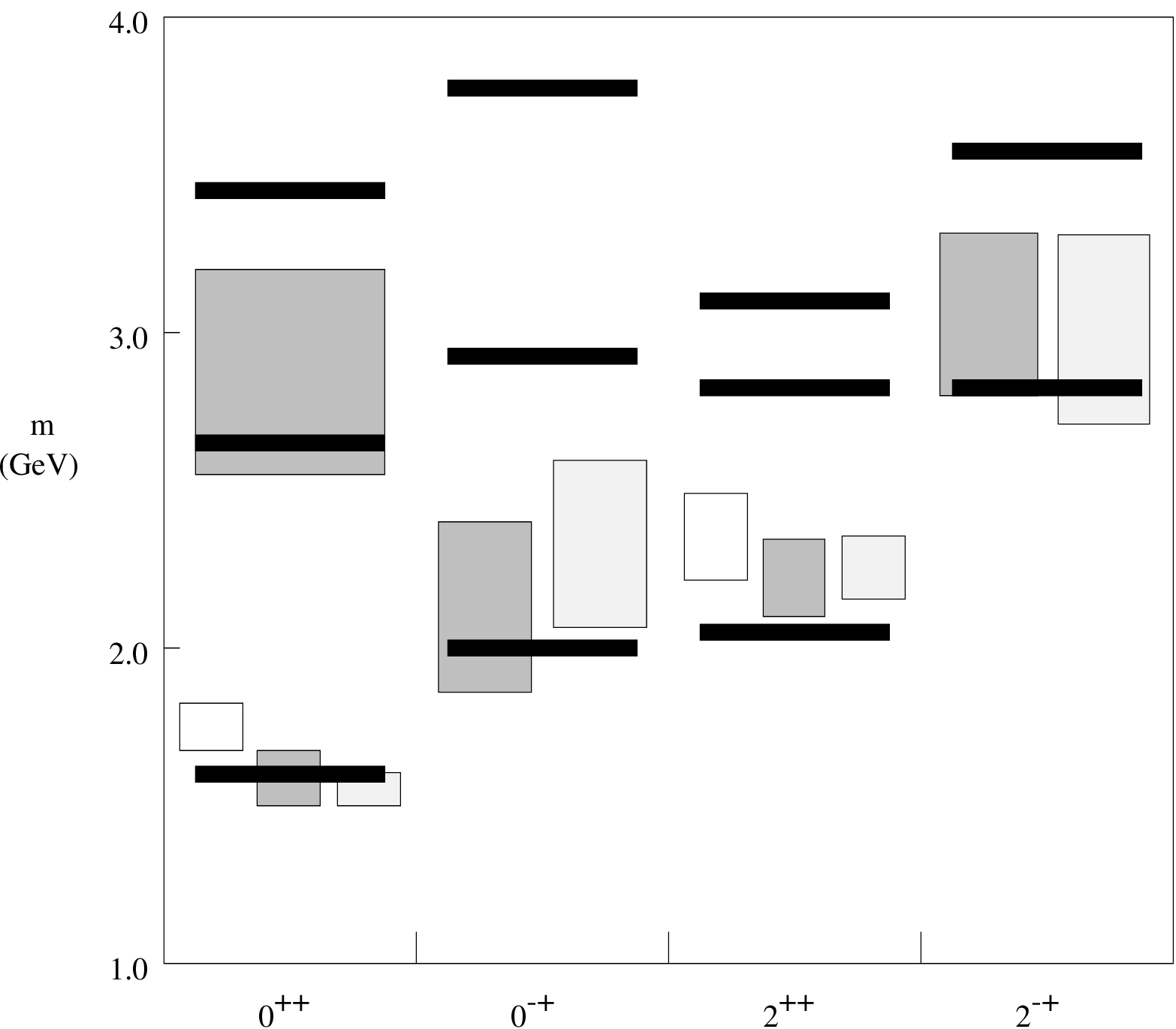}\hss}
\label{fig:spec}
\end{figure}
\end{minipage}}
\begin{center}
  {\small Fig.~1. Lattice Gauge and Model Glueball Spectra. The black bars
  are the
  results of the model calculation. Lattice results are indicated by open
  \cite{GF112},
  dark \cite{T}, and light \cite{UKQCD} boxes.}
\end{center}

Future, more comprehensive, studies may dictate
   phenomenological potentials beyond the
two-body form used here. For example, the phenomenological $^3P_0$ decay
vertex cannot be obtained from the density-density confining potential.
Reconciling the naive quark model phenomenology with the model
presented here should prove very instructive.
Another issue is that the dynamical breaking of chiral symmetry
 should lead to a massless pion solution. Recall, however,
  that
the BCS vacuum is not a true eigenstate of  the Hamiltonian. Therefore
 when diagonalizing the Hamiltonian in a truncated Fock space a
 chiral pion solution may not necessarily appear.
 However, in the random phase approximation which builds upon the BCS vacuum,
  the Nambu-Goldstone realization of chiral symmetry is preserved.
 In particular the Gell-Mann--Oaks--Renner\cite{GMOR} relation
 \begin{equation}
 f_\pi^2 m_\pi^2 = -2m_q \langle \bar q q \rangle
 \end{equation}
 follows from Thouless' theorem~\cite{TH}  applied to the chiral charge
operator,
 $Q_5 = \int d{\bf x} \Psi^{\dag}({\bf x}) \gamma_5 \Psi({\bf x}) $
 via the expression

 \begin{equation}
  2 \sum_n |\langle n| Q_5 | \Omega \rangle_{RPA}|^2(E_n - E_0)_{RPA}
  = \langle \Omega | [Q_5, [Q_5, H]] | \Omega \rangle
  \end{equation}

\noindent
Examining the
interplay of these issues will be instructive, especially as it has bearing
on the rather mysterious nature of the hyperfine splitting and the ultimate
utility of the potential quark model.

  Future work will also focus on baryon and hybrid structure.
 Towards this end, we have performed initial,  but preliminary,  calculations
in the quark sector  finding  $m_q \sim 180$ MeV and
$\langle \bar q q \rangle \sim -(100\,\mbox{Mev})^3$
in agreement with Refs.~\cite{O,AD}. While $m_q$ is in rough agreement with
phenomenology
the low condensate value may be
 due to truncation of $H_{phen}$ to two-body form. We plan to extend this to
 higher
terms and also incorporate
 the above mentioned many-body treatments.  Of special
interest will be an ambitious multi-particle/multi-hole
diagonalization which will involve higher quark Fock state
components.  In particular the importance of such states as
$|qqq (q \bar q)\rangle$
 for the proton will directly address the role of sea quarks and hidden
flavor.  Related to this is the insight this model provides concerning the
proton spin.

	In summary, we have presented a unified, comprehensive
approach to hadron structure based on non-perturbative relativistic
field theory and the QCD Hamiltonian.  The model provides both the
appealing physical insight associated with the phenomenologically
successful quark model and a consistent unified framework
 for studying issues such as chiral pions and quark-glue mixing.
  With
the advent of CEBAF a challenging opportunity is at hand to confront
new precision data and to thoroughly investigate a wide variety of
issues in hadronic physics.

	Financial support from the U. S. Department of Energy Grants
DE-FG05-88ER40461 and DE-FG05-90ER40589 and Cray-Y-MP time
from the North Carolina Supercomputer is acknowledged.  The first
two authors are grateful to Nathan Isgur, Ted Barnes, and Mike Teper
for useful discussions.

\newpage

\figure{FIG.~1. Lattice gauge and model glueball spectra. }

\end{document}